\documentclass[preprint,12pt]{elsarticle}%
\usepackage{amssymb}
\usepackage{amsfonts}
\usepackage{amsmath}
\usepackage{graphicx}%
\providecommand{\U}[1]{\protect\rule{.1in}{.1in}}
\journal{journal}

\begin{document}
%
\begin{frontmatter}%


%

\title{Adhesive rough contacts near complete contact}%

%

\author{M. Ciavarella}%
%

\address
{Politecnico di BARI. Center of Excellence in Computational Mechanics. Viale Gentile 182, 70126 Bari. Mciava@poliba.it}%
%

\begin{abstract}%

Recently, there has been some debate over the effect of adhesion on the
contact of rough surfaces. Classical asperity theories predict, in agreement
with experimental observations, that adhesion is always destroyed by roughness
except if the amplitude of the same is extremely small, and the materials are
particularly soft. This happens for all fractal dimensions. However, these
theories are limited due to the geometrical simplification, which may be
particularly strong in conditions near full contact. We introduce a simple
model for adhesion, which aims at being rigorous near full contact, where we
postulate there are only small isolated gaps between the two bodies. The gaps
can be considered as "pressurized cracks" by using Ken Johnson's idea of
searching a corrective solution to the full contact solution. The solution is
an extension of the adhesive-less solution proposed recently by Xu, Jackson,
and Marghitu (XJM model)\ (2014). This process seems to confirm recent
theories using the JKR theory, namely that the effect of adhesion depends
critically on the fractal dimension. For $D<2.5$, the case which includes the
vast majority of natural surfaces, there is an expected strong effect of
adhesion. Only for large fractal dimensions, $D>2.5$, seems for large enough
magnifications that a full fractal roughness completely destroys adhesion.
These results are partly paradoxical since strong adhesion is not observed in
nature except in special cases. \ A possible way out of the paradox may be
that the conclusion is relevant for the near full contact regime, where the
strong role of flaws at the interfaces, and of gaps full of contaminant,
trapped air or liquid in pressure, needs to be further explored. If conditions
near full contact are not achieved on loading, probably the conclusions of
classical asperity theories may be confirmed. %

\end{abstract}%
%

\begin{keyword}%

Roughness, Adhesion, Fuller and Tabor's theory, fractals%

\end{keyword}%
%

\end{frontmatter}%



\section{\bigskip Introduction}

Adhesion between elastic bodies was relatively unexplored until the last few
decades, and this is reflected in the very marginal role it has in the
otherwise very comprehensive book of K.L. Johnson (1985), despite Johnson
himself is one of the authors of one of the most important papers on adhesion
(on adhesion of elastic spheres, the JKR theory, Johnson et al., 1971, which
has over 5000 citations at present). This is obviously because until
sufficiently accurate and high-resolution technique were developed, adhesion
was hard to measure, because roughness, it was commonly observed and
explained, destroys the otherwise very strong field of attraction between
bodies, which should in principle make them stuck to each other at the
theoretical strength of the material. JKR theory itself was developed having
in mind the special cases where adhesion can indeed be measured at the
macroscopic scales, using very soft materials like rubber and gelatin spheres,
clean and with extremely smooth surfaces. Today, there is however interest in
both scientific and technological areas also at small scale, where very smooth
surfaces for example in information storage devices result in adhesive forces
playing a more crucial role than in more conventional tribological
applications. On the other hand, when people have started to study adhesion in
Geckos, which adhere to just about any surface, being it wet or dry, smooth or
rough, hard or soft, with a number of additional extraordinary features
(self-cleaning, mechanical switching), interest is emerging on how to
reproduce these capabilities in "gecko inspired synthetic adhesives". The
interest stems on the fact that adhesion cannot be produced on hard rough
surfaces, and therefore only the strikingly complex hierarchical structure of
the gecko attachment can produce the macroscopic values of load that Gecko can sustain.

The hierarchical structure of the gecko attachment (about three million
microscale hairs (setae) which in turn each branch off into several hundreds
of nanoscale spatula, totalling perhaps a billion spatula) makes one wonder
why the multiscale nature of surface roughness also could not show an effect
of adhesion enhancement. Indeed, at least one model of adhesion of solid
bodies (that of Persson and Tosatti, 2001, PT in the following), does show
adhesion persistence and even enhancement. There seems to be a qualitative
difference for surfaces with fractal dimensions below $2.5$, which turns out
to be the case in most if not the totality of the commonly observed rough
surfaces (Persson, 2014). In general, it is hard to measure strong adhesion,
despite the van der Waals interactions in principle are orders of magnitude
larger than atmospheric pressure --- this \textquotedblleft adhesion
paradox\textquotedblright\ (Pastewka and Robbins, 2014, Persson \textit{et
al}., 2005) has been linked to surface roughness, but the explanations of the
paradox have been different, the latest very interesting one being due to
Pastewka and Robbins (2014), which is a very promising parameter-free theory
that shows how adhesion changes contact area and when surfaces are sticky ---
but mostly in a regime near small contact areas. Pastewka and Robbins (2014)
conclude that "\textit{For most materials, the internal cohesive interactions
that determine elastic stiffness are stronger than adhesive interactions, and
surfaces will only stick when they are extremely smooth. Tape, geckos, and
other adhesives stick because the effect of internal bonds is diminished to
make them anomalously compliant"}. This conclusion seems in qualitative
agreement with the classical asperity theory, except that Pastewka and Robbins
use in their model quantities related to slopes and not to heigths and
therefore are in quantitative disagreement.

Persson (2002a, 2002b) introduced more elaborate version of the theory, which
solves the partial contact problem also, and the coupling of the two aspects
(effective energy due to roughness in full contact, and its use in a partial
contact with a diffusion model) makes the limit behaviour for very short
wavelengths difficult to capture, and motivated us to search a possibly
simpler, more traditional picture.

The traditional asperity model of Fuller and Tabor (1975), today is not
considered to be adequate because of its many assumptions on geometry and
absence of interaction, showed that adhesion and pull-off force is reduced
very easily at macroscopic scale by roughness. Even extremely tiny amounts of
roughness, of the order of the pull-off distance for the highest asperities in
contact, make the pull-off force orders of magnitude lower than the nominal
value. FT theory seemed to be in good agreement with the experiments, within
the limits of their accuracy. The only case where it seemed contradicted by
some experimental evidence, was in some measurements of adhesion in highly
viscoelastic solids (Fuller and Roberts, 1981, Briggs and Briscoe 1977). These
experiments indeed showed an enhancement of adhesion with roughness, which was
not expected in the pure elastic FT model. More recent evidence comes from the
cleverly designed experiments using a two-scale axisymmetric problem with
roughness between gelatin and perspex flat rough plates, by Guduru and his
group (Guduru (2007), Guduru \textit{et al} (2007), Waters \textit{et al}
(2009)). They showed clearly that an elastic JKR analysis explains the strong
increment of pull-off forces observed (an order of magnitude increase), and
that this comes with irreversible energy dissipated in many jumps of the
force-area curve.

---

In this paper, we shall try therefore a new model for a rough surface,
completely different from either asperity models, and PT model (or Persson,
2002a, 2002b). The model is based on the very simple idea Johnson used several
times in analyzing contacts near full contact, and which in turn could be
attributed to Bueckner (1958): namely, that the gaps in an otherwise full
contact are cracks that cannot sustain finite stress intensity factors in the
case of pure mechanical contact without adhesion, or that can sustain the
appropriate stress intensity factor corresponding to the toughness $K_{Ic}$(in
terms of surface energy, $G_{c}=K_{Ic}^{2}/E$), in the case of adhesion.
Further, it was used more recently by Xu \textit{et al.} (2014) (XJM theory)
for a random rough surface near full contact but without adhesion, whose model
in fact inspired the present extension to the case with adhesion.

\section{Preliminary remarks on a single sinusoid contact}

Before embarking into the full rough surface case, it is crucially important
to understand qualitatively the mechanics of adhesion near full contact. The
best strategy is to start from the relatively simple behavior of a single
sinusoidal contact, as studied quantitatively by Ken Johnson under the JKR
regime assumption (Johnson, 1995). Taking therefore a sinusoid (in either 1D
or full 2D) with $\lambda$ wavelength, $h$ amplitude, and considering the
limit case without adhesion $p^{\ast}=\pi E\frac{h}{\lambda}$ is the
compressive mean pressure to flatten the sinusoid and achieve full contact,
the adhesive case follows curves of area-load described in Fig.1, where we
have considered the case of a 1D profile for simplicity because it is fully
analytical, whereas probably the 2D case cannot be solved in closed form,
except near full contact and near pull-off. Starting from the case of "low
adhesion", $\alpha<0.6$, we can describe the behavior during loading as
follows. The curve has two extremes, a minimum and a maximum: under zero load,
the contact jumps into a state of contact given by the intersection of the
curve with the load axis. Upon further increase of the load, it follows the
stable curve, until it jumps into full contact at the maximum. At this point
the strength is theoretically infinite (more precisely, the theoretical stress
of the material, which is very high) unless we postulate the existence of some
flaw of trapped air, as Johnson suggests, and which gives a bounded tensile
pressure for returning on the curve at the maximum. Upon further unloading,
the curve is followed stably until the minimum is reached (therefore we have a
new part of the curve that is now stable, that under negative loads), where
pull-off, or jump out-of-contact, is obtained. For a "critical" value of
adhesion, which depends on modulus of the material as well as the two length
scales in the problem, the surfaces will spontaneously snap into contact at
zero load. This occurs for Johnson's parameter
\begin{equation}
\alpha=\sqrt{\frac{2\lambda\Delta\gamma}{\pi^{2}h^{2}E}}>0.6\label{alfa}%
\end{equation}
where $\Delta\gamma$ is the surface energy. What matters in particular to the
present investigation is that the original contact curve (that without
adhesion) changes sharply shape when adhesion is introduced, since the
negative pressure region appears which is crucial to understand pull-off
loads, and the transition towards infinite tension is also introduced, rather
than having full contact at specific value of (compressive) pressure. However,
for what concerns the condition of jump into full contact, this is essentially
given by the maximum of the curve which is a perturbation of the contact
solution --- the curve for negative tension and pull-off requires a detailed
analysis of the regime of low contact, and we believe for this part the model
of Fuller and Tabor (1975) is a good starting point, as the Hertzian JKR
solution is a good starting point to study the sinusoidal indenter. Notice
finally that for $\alpha>0.6$ we simply have that the pull-off force will be
load dependent.

\bigskip

\begin{center}
\frame{\includegraphics[width=100mm,
keepaspectratio=true]{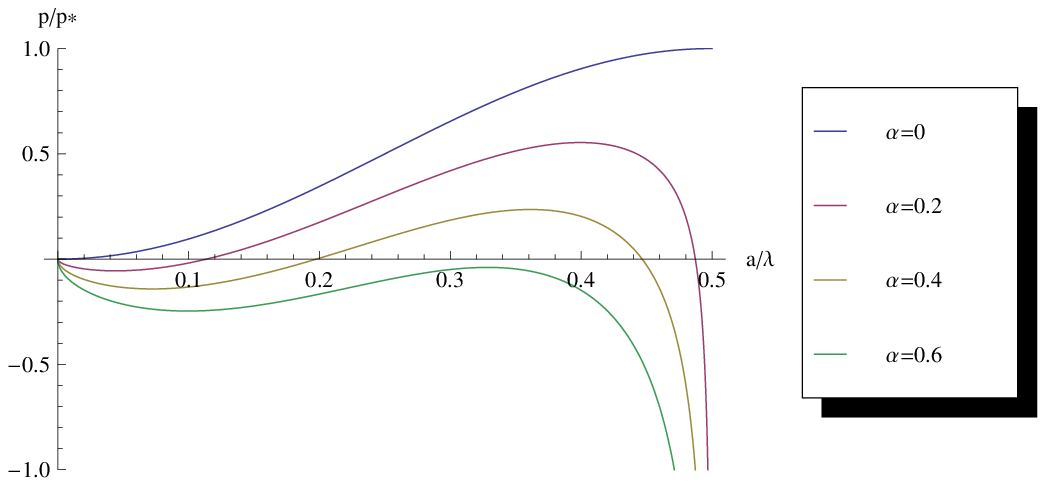}}

\vspace{2mm} \textbf{Figure 1. The relationship between }$p/p^{\ast}%
$\textbf{\ and contact area ratio }$a/\lambda$\textbf{\ for Johnson's JKR
solution of the single 1D sinusoidal adhesive contact problem.\ The change
from the pure contact case }$\alpha=0$, \textbf{to adhesive case }%
$\alpha=0.2,0.4,0.6$.
\end{center}

\vspace{5mm}

\section{The model}

In the classical random process theory, the pressure to cause full contact,
$p_{fc}$, is a random variable, whose variance is easily related in terms of
the power spectrum density (PSD) of the profile or of the surface (see Manners
and Greenwood, 2006, Persson, 2001, Persson and Tosatti, 2001)
\begin{equation}
V\left(  \zeta\right)  =\left\langle p_{fc}^{2}\right\rangle =\frac{1}%
{4}E^{\ast2}\sigma_{m}^{2}=\frac{1}{4}E^{\ast2}m_{2} \label{variance}%
\end{equation}
where $\sigma_{m}^{2}=\left\langle \left(  \frac{dz}{dx}\right)
^{2}\right\rangle =m_{2}$ is the variance of the slopes. $E^{\ast}$ here is
the combined plane strain modulus of the contact materials. The distribution
of pressures in full contact is also a Gaussian distribution, namely%
\begin{equation}
P\left(  p,\overline{p},V\right)  =\frac{1}{\sqrt{2\pi V}}\exp\left[  -\left(
\frac{\left(  p-\overline{p}\right)  ^{2}}{2V}\right)  \right]
\label{pfullcontact}%
\end{equation}
This means that, strictly speaking, there is always a tail of negative
(tensile) pressures, and indeed Persson's solution for the adhesionless
contact problem simply truncates this distribution by subtracting from it a
specular distribution, with negative mean pressure, obtaining in fact a
Rayleigh distribution.

For typical self-affine surfaces, eliminating the so called "roll-off"
wavelength, the power spectrum is a power law above a certain long wavelength
cut-off (wavenumber $q_{0}$)%
\begin{equation}
C\left(  q\right)  =\left\{
\begin{array}
[c]{cc}%
0 & \text{for }q<q_{0}\\
\frac{H}{2\pi}\left(  \frac{h_{0}}{\lambda_{0}}\right)  ^{2}\left(  \frac
{q}{q_{0}}\right)  ^{-2\left(  H+1\right)  } & \text{for }q>q_{0}%
\end{array}
\right.
\end{equation}
For any finite short wavelength cutoff $q_{1}=\zeta q_{0}$, the moments are
bounded, but in the limit, $m_{0}$ only is bounded (the variance of the
heights), whereas $m_{2}$ diverges as well as the other higher order ones. For
this reason, it is already well known that full contact cannot occur for any
finite pressure, in the fractal limit, as it is for determinist fractal
profiles like those defined with the Weierstrass series in Ciavarella et al.
(2000). In the random case, Persson's solution, which is approximate but
qualitatively correct (Wolf et al, 2014) shows that the contact area is only
complete for infinite applied compression.

\bigskip

According to Bueckner's principle (1958), as used by Johnson (1995) and XJM
(Xu \textit{et al.} 2014), we need to look carefully at the tensile stresses
of the full contact pressure solution, which are applied by superposition to
the gaps area, and compute the stress intensity factors. Considering this
contact pressure solution as a random process, it is clear that its moments
will correspond to the original surface moments, shifted by a factor 2 and
multiplied by $\frac{1}{4}E^{\ast2}$%
\begin{equation}
m_{0}^{p}=\left\langle p_{fc}^{2}\right\rangle =\frac{1}{4}E^{\ast2}%
m_{2};\qquad m_{2}^{p}=\frac{1}{4}E^{\ast2}m_{4};\qquad m_{4}^{p}=\frac{1}%
{4}E^{\ast2}m_{6}%
\end{equation}

According to Taylor series expansion therefore, as Manners and Greenwood
(2006), and XJM (Xu \textit{et al.} 2014) also suggest, the tensile part of
the full contact solution can be approximated with quadratic equations near
the pressure tensile summits, similarly to what is done for asperity theories
for the real geometry of the surface. This generates a model of isolated
"non-contact" areas. The XJM model (Xu \textit{et al.} 2014) shows that this
leads to the following:-

\begin{itemize}
\item the individual tensile area is
\begin{equation}
A_{t}=\pi a_{t}^{2}=2\pi R_{p}\left(  p-\overline{p}\right)
\label{area_tension}%
\end{equation}
where $R_{p}$ is the radius of the asperity full contact pressure
distribution, and $\left(  p-\overline{p}\right)  $ is the pressure on the
asperity (the full contact solution in the tensile regions, with a change in
sign due to the Bueckner's superposition (1958));

\item the non-contact area can be exactly shown, according to Bueckner's
principle, to correspond to a pressurized crack.

\item For the pure contact case, the stress intensity factor (SIF) has to be
zero along the boundary, axysimmetric by assumption (consideration of
elliptical form do not change the results significantly)%
\begin{equation}
K_{I}=\frac{2\sqrt{c_{t}}}{\sqrt{\pi}}\left(  p-\overline{p}\right)  \left[
1-\frac{2}{3}\frac{c_{t}^{2}}{a_{t}^{2}}\right]  =0
\end{equation}
leading to the conclusion that the non-contact area is larger than the tensile
stress area, and in particular, it is $3/2$ in size of the original tensile
area,
\begin{equation}
c_{t}/a_{t}=\sqrt{3/2}\label{ct_noadhesion}%
\end{equation}
Notice however that the condition is also satisfied trivially by the solution
$c_{t}=0$ --- since clearly the fact that the size of the gap goes to zero is
also a solution is problematic for studying that limit. The point is we also
have another condition and not just LEFM: namely, that there cannot be any
tension, which means actually the strength is zero. If we consider adhesion,
there can be tension, up to the theoretical strength.

\item For the case where there is in fact surface energy, the condition
becomes more elaborate
\begin{equation}
K_{I}=\frac{2\sqrt{c_{t}}}{\sqrt{\pi}}\left(  p-\overline{p}\right)  \left[
1-\frac{2}{3}\frac{c_{t}^{2}}{a_{t}^{2}}\right]  =K_{Ic} \label{KIc}%
\end{equation}
and for example if $\frac{c_{t}^{2}}{a_{t}^{2}}=1$, we would have an open
crack $K_{I}=\frac{2\sqrt{c_{t}}}{\sqrt{\pi}}\left(  p-\overline{p}\right)  >0
$ that would tend to propagate. We expect naturally $\frac{c_{t}^{2}}%
{a_{t}^{2}}<3/2$ as an effect of adhesion.

\item There cannot be solutions below the size where we need to take into
account the transition towards a strength criterion --- $\left(
p-\overline{p}\right)  =\sigma_{th}$. However, this is of concern only if
there is a solution of full contact with finite pressure, and in any case, the
suggestion of Johnson to consider at this point the presence of trapped air
and we shall return later on this point.
\end{itemize}

Substituting (\ref{area_tension}), into (\ref{ct_noadhesion}), we have an
equation for $c_{t}$
\begin{equation}
\frac{2\sqrt{c_{t}}}{\sqrt{\pi}}\left(  p-\overline{p}\right)  \left[
1-\frac{2}{3}\frac{c_{t}^{2}}{2R_{p}\left(  p-\overline{p}\right)  }\right]
=K_{Ic} \label{KIc2}%
\end{equation}
It is clear that we cannot solve this equation easily in a rigorous sense,
since $\left(  p-\overline{p}\right)  $ is a random variable. This equation
(\ref{KIc2}) is an implicit equation which defines
\begin{equation}
c_{t}=g\left(  p-\overline{p},R_{p},K_{Ic}\right)
\end{equation}
and acts like the basic function defining the local area as a function of the
"separation" $p-\overline{p}$ in the equivalent asperity model created by the
tensile full contact pressure "surface".

\subsection{The adhesion-less case}

It is useful to derive separately this case, as done by Xu et al (2014). From
(\ref{KIc2}), with $K_{Ic}=0$%
\begin{equation}
c_{t}^{2}=3R_{p}\left(  p-\overline{p}\right)  \label{KIc4}%
\end{equation}
There is no minimum tensile tension that can be sustained (unlike with
adhesion) and the integration proceeds simply as suggested by eqt.38 of Xu et
al (2014)%
\begin{equation}
\left(  \frac{A_{nc}\left(  \overline{p}\right)  }{A_{0}}\right)  =3\pi
R_{p}\eta\int_{\overline{p}}^{\infty}\left(  p-\overline{p}\right)
\Phi\left(  p\right)  dp
\end{equation}
where $\eta$ is the asperity density of the full contact pressure surface,
$\Phi\left(  p\right)  $ is the distribution of the pressures summits in this
surface, and hence it can be solved easily. Notice the integral is simply of
the same mathematical form as in the standard Greenwood and Williamson's
(1966) theory, where mean separation is replaced by mean pressure, and the
geometrical surface is replaced by the pressure surface. In the present form,
we obtain%
\begin{equation}
\frac{A_{nc}\left(  \overline{p}\right)  }{A_{0}}=3\pi R_{p}\eta\sqrt
{V}\left(  \frac{1}{\sqrt{2\pi}}\exp\left[  -\left(  \frac{\overline{p}^{2}%
}{2V}\right)  \right]  -\frac{1}{2}\frac{\overline{p}}{\sqrt{V}}Erfc\left(
\frac{\overline{p}}{\sqrt{2V}}\right)  \right)
\end{equation}

Further, at sufficiently large magnifications, $\sqrt{V}\sim\zeta^{1-H}$, and
$R_{p}\simeq\frac{0.375}{E^{\ast}}\sqrt{2\pi/m_{6}}\sim\zeta^{-\left(
3-H\right)  }$, while also the density of asperities $\eta=\frac{m_{6}}%
{6\sqrt{\pi}m_{4}}\sim\frac{\zeta^{\left(  6-2H\right)  }}{\zeta^{\left(
4-2H\right)  }}=\zeta^{2}$, which gives $R_{p}\eta\sqrt{V}\rightarrow
\zeta^{-\left(  3-H\right)  }\zeta^{2}\zeta^{1-H}=\delta$ where $\delta$ is a
prefactor of the order 1 (the exact prefactor to make the area of contact zero
at zero pressure is $\delta=\sqrt{2\pi}\sim2.5$, but this is not the correct
value at large pressures where we are concentrating our efforts)%
\begin{equation}
\frac{A_{nc}\left(  \overline{p}\right)  }{\delta A_{0}}\rightarrow\frac
{1}{\sqrt{2\pi}}\exp\left[  -\left(  \frac{\overline{p}^{2}}{2V}\right)
\right]  -\frac{1}{2}\frac{\overline{p}}{\sqrt{V}}Erfc\left(  \frac
{\overline{p}}{\sqrt{2V}}\right)
\end{equation}
This suggests that if we want to keep a constant value of given area of gap,
upon increasing magnification, we need to keep $\frac{\overline{p}}{\sqrt{2V}%
}$ constant, i.e. increase the mean pressure without limit. This is the well
know behavior of pure contact problem, and it is confirmed here.

\subsection{Energy balance equation at boundary of gaps}

It is important to discuss in details equation (\ref{KIc2}), as it governs the
basic behavior of the gaps during the loading process. First of all, it is
easier to manipulate it in terms of the tension on the "pressurized cracks",
as (notice that $R_{p}$ being the radius of the pressure surface, has
dimensions $m^{4}/N$), by rewriting it as
\begin{equation}
\left(  p-\overline{p}\right)  =\frac{2}{3}\frac{c_{t}^{2}}{2R_{p}}%
+\frac{\sqrt{\pi}}{2\sqrt{c_{t}}}K_{Ic} \label{KIc3}%
\end{equation}
where the second term cancels out obviously in the adhesion-less case, but
also with adhesion at very large $c_{t}$. A few curves could be plotted to
show the general trend of introducing a minimum which clearly corresponds to
the maximum in the case in Johnson's sinusoidal case (see Fig.1) near the
(unstable) transition to full contact. We prefer however to arrive at a
cleaner plot, which will be in Fig.2, to include $K_{Ic}$ in a unique curve.
Notice that as we increase the mean compression in the contact, the actual
value of tension in the gaps decreases --- therefore, the loading progresses
here by reducing the pressure on the y-axis. The minimum occurs at
$\frac{\partial}{\partial c_{t}}\left(  p-\overline{p}\right)  =\frac{2}%
{3}\frac{c_{t}}{R_{p}}-1/2\frac{\sqrt{\pi}}{2c_{t}^{3/2}}K_{Ic}=0$, which
gives
\begin{equation}
c_{t\min}=\left(  \frac{3}{8}\sqrt{\pi}\right)  ^{2/5}\left(  R_{p}%
K_{Ic}\right)  ^{2/5} \label{cmin}%
\end{equation}
in which case, for $\psi=\left(  \frac{1}{3}\left(  \frac{3}{8}\sqrt{\pi
}\right)  ^{4/5}+\frac{\sqrt{\pi}}{2\left(  \frac{3}{8}\sqrt{\pi}\right)
^{1/5}}\right)  =\allowbreak1.\,\allowbreak2021$
\begin{equation}
\left(  p-\overline{p}\right)  _{\min}=p_{0}=\psi\left(  \frac{K_{Ic}^{4}%
}{R_{p}}\right)  ^{1/5} \label{pmin}%
\end{equation}
$\,$ where $R_{p}$ is a radius of pressure "asperity". Hence, we can rewrite
(\ref{KIc3}) as%
\begin{equation}
\frac{\left(  p-\overline{p}\right)  }{p_{0}}=\frac{2}{3}\frac{1}{\psi}%
\frac{c_{t}^{2}}{2R_{p}}\left(  \frac{R_{p}}{K_{Ic}^{4}}\right)  ^{1/5}%
+\frac{1}{\psi}\frac{\sqrt{\pi}}{2\sqrt{c_{t}}}K_{Ic}\left(  \frac{R_{p}%
}{K_{Ic}^{4}}\right)  ^{1/5}. \label{eqz-intermedia}%
\end{equation}

The result comes clean defining
\begin{equation}
\widehat{p}=\frac{\left(  p-\overline{p}\right)  }{p_{0}}\text{ and }%
\widehat{c}_{t}=\frac{c_{t}}{c_{t\min}}%
\end{equation}
Hence, we can rewrite (\ref{eqz-intermedia}) as
\begin{equation}
\widehat{p}=\frac{1}{5}\left(  \widehat{c}_{t}^{2}+\frac{4}{\sqrt{\widehat
{c}_{t}}}\right)  \label{basic-clean}%
\end{equation}
and this is indeed the curve plotted in Fig. 2. In the adhesion-less case,
$\left(  p-\overline{p}\right)  _{\min}\rightarrow\infty$, and $c_{t\min
}\rightarrow\infty$ and hence the curve diverges to infinity and hence this
simple unique curve for the adhesive case represents a discontinuity and the
dimensional quantities should be plotted to see more clearly the transition
from the adhesion-less case to the adhesive one.

\begin{center}
\frame{\includegraphics[width=140mm,
keepaspectratio=true]{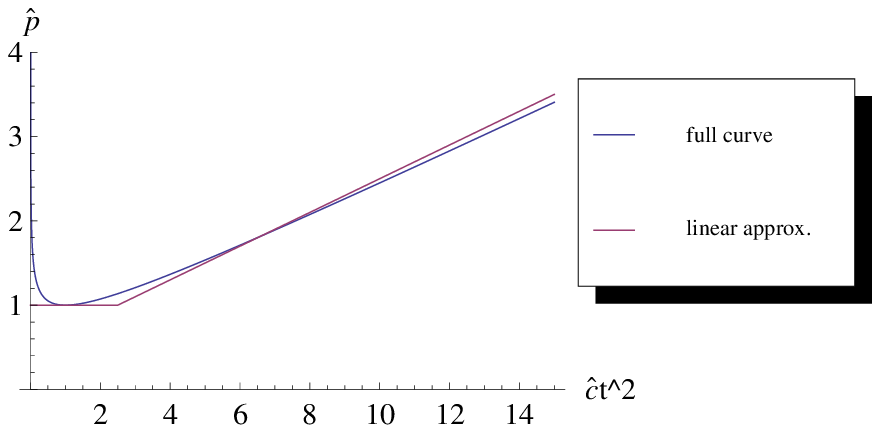}}

\vspace{2mm} \textbf{Figure 2. The relationship between peak dimensionless
tension }$\widehat{p}$ \textbf{applied on a pressurized isolated gap\ and the
size of the dimensionless gap radius squared }$\widehat{c}_{t}^{2}%
$\textbf{\thinspace, together with a linear asymptotic approximation which
turns out to be reasonably accurate also at low tensions. The minimum in the
curve, which leads to the transition to full contact, is obviously
overestimated by the approximate curve, so there will be a minor spurious
effect of increase of area of the gaps. The way the plot is constructed
doesn't permit to see the limit of pure contact (adhesive-less case), since
both }$c_{t\min}$\textbf{\ and }$p_{0}$\textbf{\ are zero in that limit, and
hence the adhesive-less curve becomes the very tail of the present one at
infinity }
\end{center}

Further, considering that we want to make estimates of the area of the gap
that remains in contact, it is clear that a very good approximate solution
could be
\begin{equation}
\widehat{p}=\left\{
\begin{array}
[c]{cc}%
\frac{1}{5}\widehat{c}_{t}^{2}+\frac{1}{2} & \text{for }\widehat{c}_{t}%
^{2}>\frac{5}{2}\\
1 & \text{for }0<\widehat{c}_{t}^{2}<\frac{5}{2}%
\end{array}
\right.  \label{p-approx}%
\end{equation}
since the other branch of the equation is unstable. The only significant
approximation this introduces, it is to overestimate the size of the gap
region radius just before jumping into contact. However, while this will only
change some prefactors by small amounts, it simplifies the study of the
problem enormously.

Now, supposing we start from a contact which doesn't jump into contact
completely and with an applied compressive load --- but that gaps area do
exist. It is clear that the (\ref{p-approx}) solution holds for each gap,
depending on the local tension arising from cancelling the tension in the full
contact pressure, and at any given mean applied load, each gap will have a
certain level of pressure and therefore a certain equilibrium size as per
(\ref{p-approx}), which may include some gaps closing. We need an integration
process to establish the total area of gaps, and therefore the complementary
remaining contact area. Also, upon loading, the mean compressive pressure
increases, and hence the tensile pressure on each gap decreases, so that more
gaps will tend to reach the condition $\widehat{p}$ $=1$. Therefore, a certain
number of gaps will close, and the others will reduce their size. To
understand if the final state is full contact or not, we should consider the
adhesion-less case for reference. In this case, approximately, Persson's
solution indicates that only $\sqrt{V}$ is the parameter ruling the contact
area size. If we increase the short wavelength content, increasing $\sqrt{V}$,
for a given contact area, we have to increase the mean pressure in proportion.
Given $\sqrt{V}$ grows unbounded, the pressure to obtain any value of contact
area (in fact, not just full contact), grows unbounded.

Repeating this reasoning here, we need to observe if, for a given condition
with adhesion, the contact area depends only on the ratio $p/\sqrt{V}$ or not.

Rewriting and inverting the equation (\ref{p-approx}) in terms of the gap
radiuses%
\begin{equation}
\widehat{c}_{t}^{2}=\left\{
\begin{array}
[c]{cc}%
5\left(  \widehat{p}-\frac{1}{2}\right)  & \text{for }\widehat{p}>1\\
0 & \text{for }\widehat{p}<1
\end{array}
\right.
\end{equation}
and in dimensional terms%
\begin{equation}
c_{t}^{2}=\left(  \frac{3}{8}\sqrt{\pi}\right)  ^{4/5}\left(  R_{p}%
K_{Ic}\right)  ^{4/5}\left\{
\begin{array}
[c]{cc}%
5\left(  \frac{\left(  p-\overline{p}\right)  }{p_{0}}-\frac{1}{2}\right)  &
\text{for }\frac{\left(  p-\overline{p}\right)  }{p_{0}}>1\\
0 & \text{for }\frac{\left(  p-\overline{p}\right)  }{p_{0}}<1
\end{array}
\right.  \label{ct-final}%
\end{equation}
or%
\begin{equation}
\pi c_{t}^{2}=3\pi R_{p}\left[  \left(  p-\overline{p}\right)  -p_{0}\right]
\text{ , for }\left(  p-\overline{p}\right)  >p_{0}%
\end{equation}

\subsection{\bigskip Integration and results}

We shall\ neglect the variation of $R_{p}$ with height, as otherwise the
integration becomes too cumbersome. We simply assume a mean value given by
random process theory, and develop an integration of the type$\ $
\begin{equation}
\frac{A_{nc}\left(  \overline{p}\right)  }{A_{0}}=\eta\int_{\overline{p}%
+p_{0}}^{\infty}\pi c_{t}^{2}\Phi\left(  p\right)  dp
\end{equation}
where the function for $c_{t}$ is now given in (\ref{ct-final}), which results
in
\begin{equation}
\frac{A_{nc}\left(  \overline{p}\right)  }{A_{0}}=3\pi R_{p}\eta\sqrt
{V}\left[  \frac{1}{\sqrt{2\pi}}\exp\left[  -\left(  \frac{\left(
\overline{p}+p_{0}\right)  ^{2}}{2V}\right)  \right]  -\frac{\overline
{p}+p_{0}}{2\sqrt{V}}Erfc\left(  \frac{\overline{p}+p_{0}}{\sqrt{2V}}\right)
\right]
\end{equation}
which agrees with the adhesiveless case when $p_{0}=0$. In this format, it is
clear that the effect of $p_{0}$ is exactly analogous to an increase of the
mean pressure. The non-contact area would tend to stay constant with
magnification $R_{p}\eta\sqrt{V}\rightarrow\delta$ now if we keep increase,
instead of the applied pressure proportionally to $\sqrt{V}\sim\zeta^{1-H}$,
only $p_{0}$, and this is simple to study. At sufficiently large
magnifications, looking at the parameter (\ref{pmin}), and considering the
usual scaling arguments on the PSD and its moments$\ $
\begin{equation}
\frac{p_{0}}{\sqrt{V}}=\frac{\psi}{\sqrt{V}}\left(  \frac{K_{Ic}^{4}}{R_{p}%
}\right)  ^{1/5}\sim\frac{m_{6}^{1/10}}{m_{2}^{1/2}}\sim\frac{\left(
\zeta^{\left(  6-2H\right)  }\right)  ^{1/10}}{\zeta^{1-H}}\sim\zeta^{\frac
{2}{5}\left(  2H-1\right)  }%
\end{equation}
where the exponent $\theta=\frac{2}{5}\left(  2H-1\right)  $ is positive for
$H>0.5$ (low fractal dimensions), when the dimensionless $\frac{p_{0}}%
{\sqrt{V}}$ increases with magnification, or negative otherwise.

Hence, for low fractal dimensions, and a given applied mean pressure
(including zero), the non-contact area tends to decrease without limit,
implying the tendency to full contact. Naturally, the tendency will be
stronger the higher $H>0.5$ i.e. the farther from the limit case of $H=0.5$.
There seems to be some connection to the conclusions and the parameters
involved in the "effective adhesion energy" in Persson and Tosatti (2001),
which leads them to suggest that adhesion persists for low fractal dimensions
$D<2.5$ (the real range of surfaces, see Persson (2014)). In their theory,
they obtain from the elastic energy associated to the deformation in balance
with surface energy of a full spectrum of frequencies, a parameter which seems
related to
\begin{equation}
Em_{1}=\frac{E}{4}h_{0}q_{0}f\left(  H\right)  \sim\zeta^{1-2H}%
\end{equation}
where $m_{1}$ is the first order moment of the PSD. This has the same
qualitative power law behaviour (besides the $\frac{2}{5}$factor), but it
should be emphasized different from -- in PT coming from the first moment of
the surface, and in our case from a combination of roots of the second and
6-th moment.

\bigskip

\begin{center}
\frame{\includegraphics[width=120mm,
keepaspectratio=true]{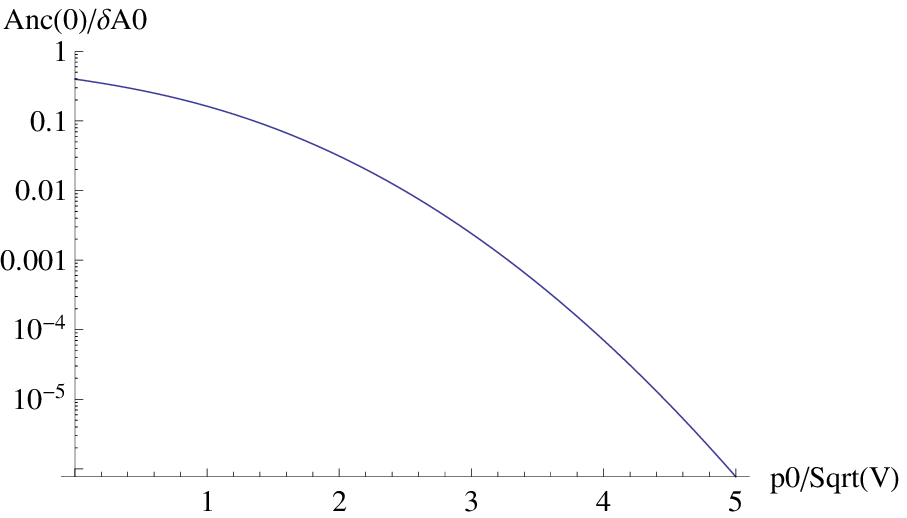}}

\vspace{2mm} \textbf{Figure 3. The area of gaps }$\frac{A_{nc}\left(
0\right)  }{A_{0}}$ \textbf{for zero applied pressure decreases rapidly with
the dimensionless adhesion pressure term }$\frac{p_{0}}{\sqrt{V}}%
$\textbf{\ and when }$\frac{p_{0}}{\sqrt{V}}\approx5\ $\textbf{we can assume
full contact holds. }
\end{center}

Fig.3 shows a plot of the non contact area at zero applied mean pressure,
$\frac{A_{nc}\left(  \overline{p}=0\right)  }{A_{0}}$, which decays very
rapidly with the dimensionless ratio $\frac{p_{0}}{\sqrt{V}}$, indicating
there is chance of spontaneous full contact for $H>0.5$, although the model is
clearly approximate in that range since we are far from full contact.

For $H<0.5$ and hence large fractal dimensions, the dimensionless ratio
$\frac{p_{0}}{\sqrt{V}}$ will decrease with magnification, and hence we return
at large magnifications to the case of pure adhesion-less contact, for which
we expect at zero load simply zero area.

The situation is clear therefore also with applied compressive forces to the
contact. Contrary to the case without adhesion, where no mean pressure is
sufficient to squeeze the contact flat, here for large Hurst exponent $H>0.5$,
the non-contact area decreases asymptotically to zero and the trends of the
zero applied pressure are confirmed, as clearly seen in Fig.4, where we assume
the power law scaling $\frac{p_{0}}{\sqrt{V}}=\zeta^{\frac{2}{5}\left(
2H-1\right)  }$ with $\frac{p_{0}}{\sqrt{V}}\left(  \zeta=1\right)  =1$.

\begin{center}
$%
\begin{array}
[c]{cc}%
\frame{\includegraphics[width=120mm,
keepaspectratio=true]{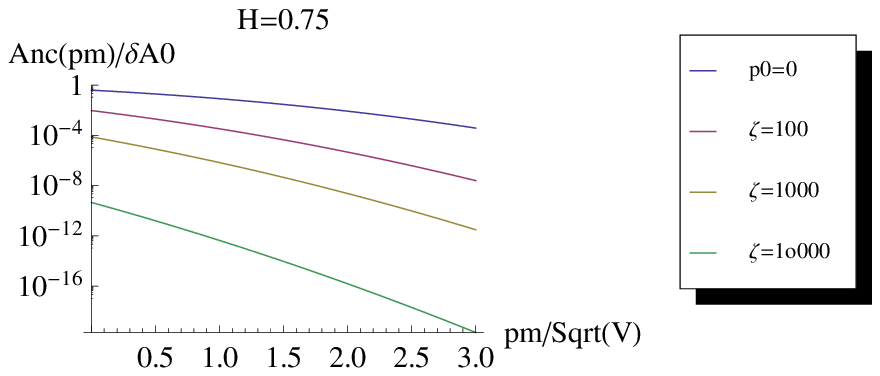}} & a\\
\frame{\includegraphics[width=120mm,
keepaspectratio=true]{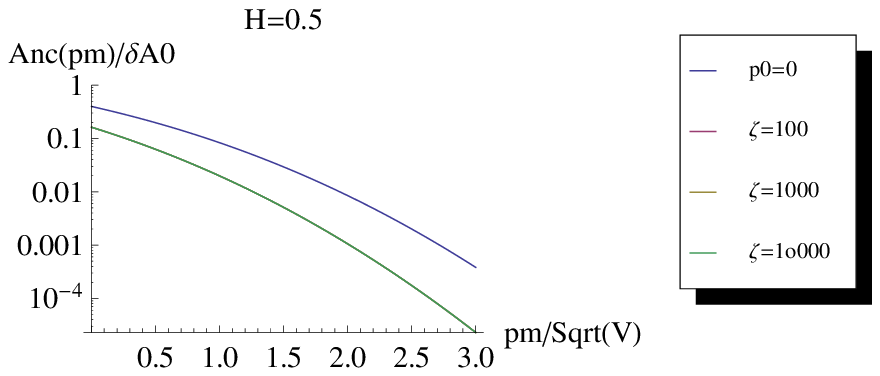}} & b\\
\frame{\includegraphics[width=120mm,
keepaspectratio=true]{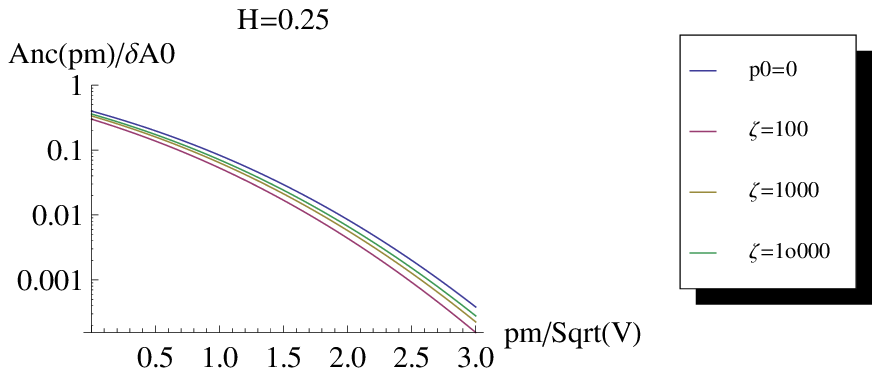}} & c
\end{array}
$

\vspace{2mm} \textbf{Figure 4. The area of gaps (complementary to the real
contact area) for different }$\mathbf{H=0.75,0.5,0.25}$\textbf{\ as a function
of the dimensionless applied compression }$\frac{\overline{p}}{\sqrt{V}}%
$\textbf{\ shows that the contact becomes practically full at finite values of
the }$\frac{\overline{p}}{\sqrt{V}}$\textbf{\ contrary to the purely
mechanical case for }$H>0.5$. \textbf{Vice versa, for }$H=0.5$\textbf{, this
tendency is very similar to the adhesive-less contact and for }$H=0.25$
\textbf{there is no significant difference with the adhesive-less case. We
assume power law scaling for }$\frac{p_{0}}{\sqrt{V}}$ \textbf{with }%
$\frac{p_{0}}{\sqrt{V}}\left(  \zeta=1\right)  =1.$\textbf{ }
\end{center}

\bigskip

\section{An alternative derivation}

We obtained a GW-like asperity full contact pressure surface model for
adhesion and more in general, in the adhesive-less version of this (obtained
putting $p_{0}=0$), Xu et al (2014) noticed the close connection to the
results with the Persson (2001) theory of contact. Xu et al (2014) introduced
varius advanced asperity models. In particular, the most advanced theories
which recognize the variation of mean radius with heigth, and for large
separations they are considered to be exact --- large separation being here,
in the equivalent pressure "surface" model, large pressures. A further attempt
is of interest, since it brings very close comparison with Persson's theory.
As recognized by the advanced theories of BGT or Carbone (2009), the contact
area can be connected the bearing area ratio $A_{b}\left(  t\right)  /A_{0}$
at large separations to the contact area, and hence, here a given separation
$d$, it depends only on a single parameter of the random process surface
($m_{0}$ is the zero-th order moment of the spectral density of the surface):
\begin{equation}
\frac{A_{b}\left(  t\right)  }{A_{0}}=\frac{1}{2}Erfc\left(  \frac{t}{\sqrt
{2}}\right)  \label{bearing-area}%
\end{equation}
where separation is made non-dimensional by the ratio $t=d/m_{0}^{1/2}$. When
translating this into the "pressure surface" model, this should give an area
of "non-contact" as a function of mean pressure $p$, made non-dimensional by
dividing it by the $0-th$ order moment of the pressures
\begin{equation}
m_{0}^{p}=\left\langle p_{fc}^{2}\right\rangle =V \label{m0}%
\end{equation}
defined above (\ref{variance}), which in fact translates onto $\frac{1}%
{4}E^{\ast}m_{2}$. Here, we recognize already some features of the Persson's
equation, contrary to the more "accurate" asperity models in Xu \textit{et
al.} (2014) which additionally depend on bandwidth parameter.

Now, in the limit when the mean pressure is zero, we know that by definition,
the non-contact area should be 1, so that the real contact area is zero.
Therefore, in order for the non-contact area to be 1 at zero separation, we
need it to be double of the bearing area (\ref{bearing-area}) at zero
separation. \ Hence, we multiply (\ref{bearing-area}) by factor 2, and use
(\ref{m0}), to get
\begin{equation}
\frac{A_{nc}\left(  p\right)  }{A_{0}}=Erfc\left(  \frac{p}{\sqrt{2m_{0}^{p}}%
}\right)  =Erfc\left(  \frac{\sqrt{2}p}{E^{\ast}\sqrt{V}}\right)
\end{equation}
Writing in terms of the complementary term, the area of contact, we get%
\begin{equation}
\frac{A_{c}\left(  p\right)  }{A_{0}}=\frac{1-A_{nc}\left(  p\right)  }{A_{0}%
}=1-Erfc\left(  \frac{\sqrt{2}p}{E^{\ast}\sqrt{V}}\right)  =\operatorname{erf}%
\left(  \frac{\sqrt{2}p}{E^{\ast}\sqrt{V}}\right)  \label{persson}%
\end{equation}
Remarkably, this is \textit{exactly} Persson's equation, in the entire range
of pressures, including the low pressure end. Therefore, the use of the
bearing area assumption is not limited by extremely large separations in the
asymptotic version of the area-separation relationship in BGT's model.

Turning back on the adhesion case, we reached the conclusion that the
integration for a GW-equivalent model was of the type $\ $
\begin{equation}
\frac{A_{nc}\left(  \overline{p}\right)  }{A_{0}}=3\pi R_{p}\eta
\int_{\overline{p}+p_{0}}^{\infty}\left[  p-\left(  \overline{p}+p_{0}\right)
\right]  \text{ }\Phi\left(  p\right)  dp
\end{equation}
and hence, this is exactly equivalent to an adhesiveless contact problem where
the mean pressure has been replaced by the sum $\left(  \overline{p}%
+p_{0}\right)  .$

Hence, in order to obtain a result consistent to adhesionless Persson's theory
in the adhesive case, we need to multiply by a factor 4/3, resulting in $\ $%
\begin{equation}
\frac{A_{nc}\left(  \overline{p}\right)  }{A_{0}}=Erfc\left(  \frac{\sqrt
{2}\left(  \overline{p}+p_{0}\right)  }{E^{\ast}\sqrt{V}}\right)
\end{equation}
resulting in
\begin{equation}
\frac{A_{c}\left(  p\right)  }{A_{0}}=\frac{1-A_{nc}\left(  p\right)  }{A_{0}%
}=1-Erfc\left(  \frac{\sqrt{2}\left(  \overline{p}+p_{0}\right)  }{E^{\ast
}\sqrt{V}}\right)  =\operatorname{erf}\left(  \frac{\sqrt{2}\left(
\overline{p}+p_{0}\right)  }{E^{\ast}\sqrt{V}}\right)
\end{equation}

Comparison with the previous theory shows that the results are qualitatively
similar. However, it is easier to plot them in terms of actual contact area,
rather than non-contact, because we forces the prefactors to be such, as in
the original Persson's theory, to produce zero contact area at zero load, in
the absence of adhesion. Fig.5 therefore plots eve the range of negative
applied load, where in some case the contact is nearly full and hence the
assumptions made of isolated gaps in the present model may be fulfilled. The
results loose sense if the contact area is small, viceversa.

\begin{center}
$%
\begin{array}
[c]{cc}%
\frame{\includegraphics[width=120mm,
keepaspectratio=true]{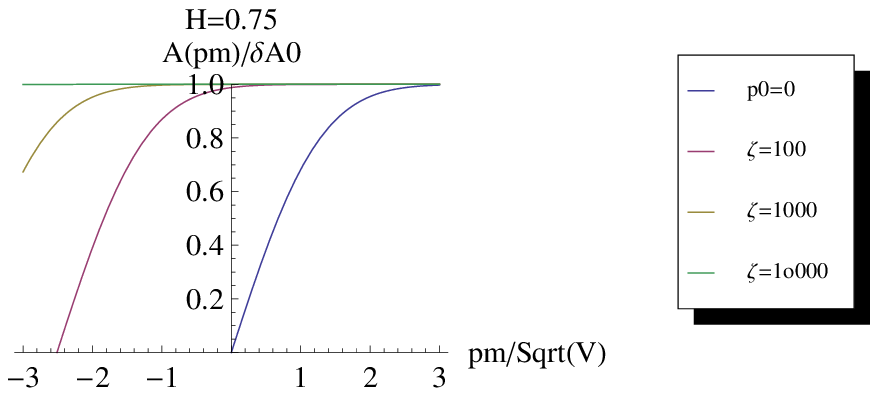}} & a\\
\frame{\includegraphics[width=120mm,
keepaspectratio=true]{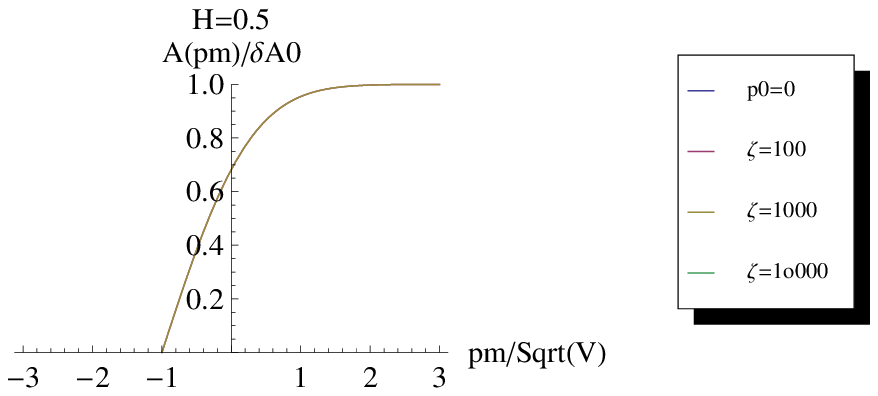}} & b\\
\frame{\includegraphics[width=120mm,
keepaspectratio=true]{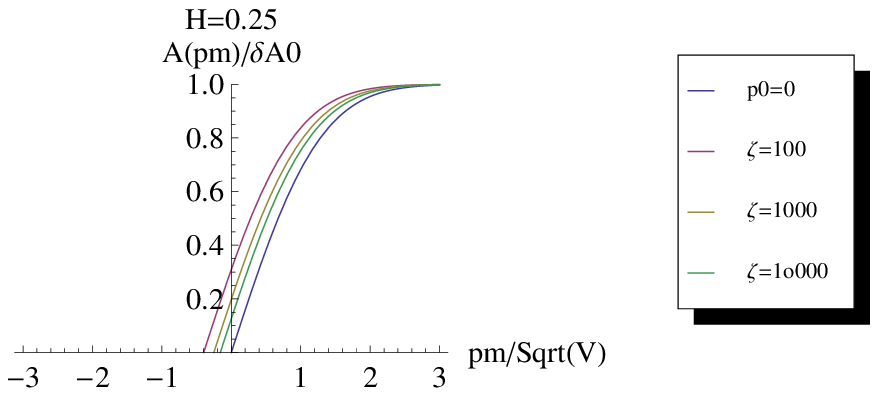}} & c
\end{array}
$

\vspace{2mm} \textbf{Figure 5. The real contact area for different
}$\mathbf{H=0.75,0.5,0.25}$\textbf{\ as a function of the dimensionless
applied compression }$\frac{\overline{p}}{\sqrt{V}}$\textbf{\ shows that the
contact becomes practically full at finite values of the }$\frac{\overline{p}%
}{\sqrt{V}}$\textbf{\ (and even zero load for }$\mathbf{H=0.75}$\textbf{,
within the limits of our model). For }$H=0.5$\textbf{, the magnification
effect is neutral, whereas for }$H=0.25$ \textbf{there is no significant
difference with the adhesive-less case, and the curves are close to the
adhesive-less case. We assume power law scaling for }$\frac{p_{0}}{\sqrt{V}}$
\textbf{with }$\frac{p_{0}}{\sqrt{V}}\left(  \zeta=1\right)  =1.$\textbf{ }
\end{center}

\subsection{Pull off?}

Let us now discuss the case of pull-off in qualitative terms. This may occur
in a range with small contact area, if either the contact doesn't proceed
spontaneously into large fractions of the nominal contact area, or if the
applied loading doesn't push toward this range. In that case contact is so
clearly isolated on asperities, than theories like Fuller and Tabor (1975)
would be approximately true.

The problem and limit of the present theory is, insteed, that it cannot deal
with the low contact areas, where gaps are interacting and certainly the bends
in the area-load curves which would appear in a set of isolated asperity
theories are not reproduced here.

The only estimate we can make is therefore to extrapolate pull-off from the
point where the contact area is expected to be zero, and this gives.%

\begin{equation}
\frac{A_{c}\left(  p\right)  }{A_{0}}=\operatorname{erf}\left(  \frac{\sqrt
{2}\left(  \overline{p}+p_{0}\right)  }{E^{\ast}\sqrt{V}}\right)
=\frac{2\sqrt{2}}{\sqrt{\pi}}\frac{\left(  \overline{p}+p_{0}\right)
}{E^{\ast}\sqrt{V}}=0
\end{equation}
which leads to a new meaning for the pressure $p_{0}$
\begin{equation}
\left(  \frac{\overline{p}}{\sqrt{V}}\right)  _{pulloff}=-\frac{p_{0}}%
{\sqrt{V}}=-\psi\left(  \frac{E^{3}G_{c}^{2}}{0.375\sqrt{2\pi}}\right)
^{1/5}\frac{m_{6}^{1/10}}{m_{2}^{1/2}}\sim-\zeta^{\frac{2}{5}\left(
2H-1\right)  }%
\end{equation}
which always increases with magnification. Notice however that in absolute
terms, $\left(  \overline{p}\right)  _{pulloff}$ is always increasing for all
parameters
\begin{equation}
\overline{p}_{pulloff}=-p_{0}\sim-m_{6}^{1/10}\sim-\zeta^{\frac{3-H}{5}}%
\end{equation}
and this result is not easy to beleive, and in opposite constrast to asperity
theories but even Pastewka and Robbins (2014) who find stick surfaces only
those that are smooth enough (in terms of surface slopes) to have the cohesive
energy in the bulk giving up against the adhesion forces at the interface.

\section{Conclusion}

A new model of adhesion has been discussed and shown to lead to very simple
and clear results: there cannot be spontaneous jump into contact for any
surface having sufficiently multiscale content, no matter its fractal
dimension. The model is devised near the full contact regime, so that the
contact consists of a set of isolated gaps whose surfaces are then loaded by
Bueckner's principle by the tensile pressures of the "linear" full contact
solution, which are approximated by parabola since, by Taylor's expansion,
they must have this form near full contact when gaps are closing, and it is
easy to write this for a Gaussian surface. The stable branch of the curve of
the gap radius vs applied pressure in the gaps are then found imposing the
stress intensity factor to be constant along the edge, and a very good
approximation turns out to be a linear law with an offset, which permits
extremely simple integration, which resemble those of asperity models, and
indeed can be considered as a "pressure asperity" model for the pressurized
gaps. In the case of no adhesion, as already noticed by Xu et al (2014), the
result turn out to be extremely similar to those of Persson's contact theory
(2001) which is a widely recognized as a good approximate solution near full
contact, and which gives us confidence the results are also very similarly
accurate where it tends to be exact in the limit of full contact (for infinite
mean pressure applied).

It is shown that a dimensionless ratio governs the contact and a pressure
$p_{0}$ can be defined which is scale dependent and includes the energy of
adhesion. This pressure has a role equivalent to the mean applied pressure in
the equation of the non-contact area and hence since it grows without limit
for low fractal dimensions, permits full contact to be achieved for those
surfaces. The conclusions cannot be complete of unloading and pull-off force,
which require further investigation.

An equation which looks like an extension of Persson's theory for contact
mechanics has been derived.

\section{References}

Briggs G A D and Briscoe B J 1977 The effect of surface topography on the
adhesion of elastic solids J. Phys. D: Appl. Phys. 10 2453--2466

Bueckner, H., 1958, The propagation of cracks and the energy of elastic
deformation. Trans of the Amer Soc of Mech Eng, 80, 1225-1230.

Ciavarella, M., et al. "Linear elastic contact of the Weierstrass profile."
Proceedings of the Royal Society of London. Series A: Mathematical, Physical
and Engineering Sciences 456.1994 (2000): 387-405.

Derjaguin, BV and Muller, VM and Toporov, Y.P., 1975, Effect of contact
deformations on the adhesion of particles, Journal of Colloid and Interface
Science, 53(2), pp. 314-326.

Fuller, K.N.G. , Roberts A.D. 1981. Rubber rolling on rough surfaces J. Phys.
D Appl. Phys., 14, pp. 221--239

Greenwood, J.A., Williamson, J.B.P., 1966. Contact of nominally flat surfaces.
Proc. R. Soc. London A295, 300--319.

Guduru, P.R. 2007. Detachment of a rigid solid from an elastic wavy surface:
theory J. Mech. Phys. Solids, 55, 473--488

Guduru, P.R. , Bull, C. 2007. Detachment of a rigid solid from an elastic wavy
surface: experiments J. Mech. Phys. Solids, 55, 473--488

Johnson, K. L., K. Kendall, and A. D. Roberts. 1971. Surface energy and the
contact of elastic solids. Proc Royal Soc London A: 324. 1558.

Johnson, K.L., 1985. Contact Mechanics. Cambridge University Press.

Johnson K.L. 1995. The adhesion of two elastic bodies with slightly wavy
surfaces, Int J Solids and Struct 32 (3-4), , pp. 423-430

Kellar. A. 2007. Gecko adhesion: structure, function, and applications. MRS
bulletin 32.06 (2007): 473-478.

\bigskip

Longuet-Higgins, MS. 1962. The statistical geometry of random surfaces, Proc.
Symp. Appl. Math.

Manners, W., Greenwood, J.A., 2006. Some observations on Persson's diffusion
theory of elastic contact. Wear 261, 600--610.

\bigskip Muller, VM and Derjaguin, BV and Toporov, Y.P., 1983, On two methods
of calculation of the force of sticking of an elastic sphere to a rigid plane,
Colloids and Surfaces, 7(3), pp. 251-259.

Pastewka, L., Robbins MO, 2014. Contact between rough surfaces and a criterion
for macroscopic adhesion, Proc Natl Acad Sci USA. 111(9): 3298--3303.

Persson, B.N.J., 2001. Theory of rubber friction and contact mechanics. J.
Chem. Phys. 115, 3840--3861.

Persson, B. N. J., Tosatti. E. 2001. he effect of surface roughness on the
adhesion of elastic solids. The Journal of Chemical Physics 115.12: 5597-5610.

Persson B N J 2002 Adhesion between an elastic body and a randomly rough hard
surface Eur. Phys. J. E 8 385;

Persson, B. N. J. 2002. Adhesion between elastic bodies with randomly rough
surfaces." Physical review letters 89.24: 245502.

Persson, B N J, Albohr, O, Tartaglino, U, Volokitin A I and Tosatti. E
\textit{\ }2005. On the nature of surface roughness with application to
contact mechanics, sealing, rubber friction and adhesion J. Phys.: Condens.
Matter 17 R1,\textit{\ http://arxiv.org/pdf/cond-mat/0502419.pdf)}

Persson, B. N. J.. 2014. On the fractal dimension of rough surfaces. Tribology
Letters 54.1, 99-106.

Xu, Y., Jackson, R.L., Marghitu, D.B., 2014. Statistical model of nearly
complete elastic rough surface contact. Int. J. Solids Struct. 51, 1075--1088.

Yastrebov, V.A., Anciaux, G., Molinari, J.F., 2014. From infinitesimal to full
contact between rough surfaces: evolution of the contact area. Int. J. Solids
Struct. Available from:
$<$%
http://arxiv.org/abs/1401.3800%
$>$%
.

Waters, J.F. Leeb, S. Guduru, P.R. 2009. Mechanics of axisymmetric wavy
surface adhesion: JKR--DMT transition solution, Int J of Solids and Struct 46
5, 1033--1042

Wolf B.D, Prodanov, N, and M\"{u}ser, MH. "Systematic analysis of Persson's
contact mechanics theory of randomly rough elastic surfaces." Journal of
Physics: Condensed Matter 26.35 (2014): 355002.

\end{document}